\documentclass[%
 reprint,
superscriptaddress,
nofootinbib,
nobibnotes,
 amsmath,amssymb,
 aps,
longbibliography
]{revtex4-2}

\usepackage{graphicx}% Include figure files
\usepackage{dcolumn}% Align table columns on decimal point
\usepackage{bm}% bold math
\usepackage[colorlinks=true,citecolor=blue,linkcolor=blue, urlcolor=cyan]{hyperref} 

\usepackage{empheq}
\usepackage{xcolor}
\usepackage[normalem]{ulem}
\usepackage[english]{babel}
\newcommand{\Tr}{\mathop{\rm Tr}\nolimits}

\begin{document}

\preprint{APS/123-QED}

%\title{Anomalous Hall effect in ballistic and hydrodynamic channels}
\title{Anomalous Hall effect in ultraclean electronic channels}

\author{K.K. Grigoryan}
\affiliation{Moscow Institute of Physics and Technology, Dolgoprudny, Russia}
\affiliation{L. D. Landau Institute for Theoretical Physics, 142432 Chernogolovka, Russia}
\author{D.S. Zohrabyan}
\affiliation{Moscow Institute of Physics and Technology, Dolgoprudny, Russia}
\affiliation{L. D. Landau Institute for Theoretical Physics, 142432 Chernogolovka, Russia}
\author{M.M. Glazov}
\affiliation{Ioffe Institute, 194021 St. Petersburg, Russia }%

\date{\today}

\begin{abstract} 
Recent technological advances allow fabricating  ultraclean two-dimensional electronic systems where  the electron mean free path due to static disorder and phonons is much larger compared to the conducting channel width. It makes possible to realize novel, ballistic and hydrodynamic, regimes of electron transport resulting in drastic modifications of the normal Hall effect. Here we develop a theory of anomalous Hall effect -- generation of the electric field transverse to the flowing current unrelated to the Lorentz force action -- in ultraclean channels with two-dimensional electron gas and demonstrate that both in ballistic and hydrodynamic regimes the anomalous Hall effect, similarly to the normal one,  strongly differs from that in the standard diffusive case. We take into account all relevant contributions to the anomalous Hall electric field and Hall voltage:  the skew scattering of electrons, side-jump, and anomalous velocity effects that appear as a result of the spin-orbit coupling. We study both ballistic and hydrodynamic transport regimes which are realized depending on the relation between the electron-electron mean free path and the channel width. The role of electron-electron interactions is analyzed. Compact analytical expressions for the anomalous Hall field and voltage are derived. Possible experimental scenarios for observation of the anomalous Hall effect in ultraclean channels are briefly discussed.
\end{abstract}

\maketitle

\section{\label{sec:intro}Introduction}

The Hall effect, namely, the production of a voltage drop and electric field across a conductor that is transverse both to an electric current flowing in the system and to an applied magnetic field perpendicular to the current~\cite{Hall:1879aa} is of prime importance in condensed matter physics. This effect underlying a plethora of magnetotransport phenomena is widely used to study fundamental properties of conducting media. The Hall effect makes it possible to determine the type of charge carriers and their density and has numerous applications~\cite{ashcroft1976solid,Ziman2001}. In semiclassical approach, the Hall effect is associated with the action of the Lorentz force on the charge carriers~\cite{Drude:1900aa}, while in the quantum-mechanical approach it is related to formation of the Landau levels and conducting channels at the sample edges being deeply connected to the topology of electronic states~\cite{PhysRevLett.45.494,PhysRevB.31.3372}.

Shortly after discovery of the normal Hall effect, it was realized that there is a contribution to the Hall voltage unrelated to the Lorentz force action and caused by the magnetization of the system~\cite{Hall:1881aa}. This effect termed as ferromagnetic or anomalous Hall effect attracts a lot of attention nowadays~\cite{RevModPhys.82.1539}. The anomalous Hall effect is related to the spin-orbit interaction and it is highly sensitive to the details of the band structure, electron scattering, and transport regimes that makes its theoretical description quite involved~\cite{PhysRev.95.1154,SMIT1955877,Adams:1959aa,gy61,abakumov72,nozieresAHE,PhysRevB.75.045315,Sinitsyn_2007,Ado_2015,PhysRevLett.123.126603}. There are three key mechanisms of the effect: the anomalous velocity acquired by spin-polarized electrons in the presence of electric field, side-jumps or shifts of  electronic wavepackets at scattering, and asymmetric or skew scattering~\cite{Sinitsyn_2007}. The mechanisms of the anomalous Hall effect are intimately connected to those of the spin and valley Hall effects~\cite{dyakonov71,dyakonov71a,hirsch99,sinova04,dimitrova:245327,chalaev:245318,Xiao:2012cr,2020arXiv200405091G,Glazov2020b}. In diffusive systems there are important cancellations of the anomalous velocity and a part of the side-jump contributions to the spin, valley, and anomalous Hall effects related to the exact balance of the electric current generation and dissipation processes in the \emph{dc} transport~\cite{dyakonov_book,2020arXiv200405091G}.

Quite recently, a class of ultraclean two-dimensional electronic systems has emerged where the electron mean free path limited by the disorder or phonon scattering is comparable or even exceeds the width of the conducting channel providing novel -- ballistic or hydrodynamic -- transport regimes~\cite{PhysRevB.51.13389,Bandurin1055,Crossno:2016aa,Moll1061,Krishna-Kumar:2017wn,Gusev:2018tg,PhysRevLett.128.136801}. In these systems the electron momentum is mainly lost at the diffusive scattering at the sample edges while dissipation in the `bulk' of the channel is suppressed. It gives rise to a number of nontrivial transport and magnetotransport phenomena~\cite{PhysRevLett.106.256804,PhysRevB.92.165433,Levitov:2016aa,PhysRevLett.117.166601,Narozhny:2017vc,PhysRevB.100.115401,Narozhny:2022ud}, some of those predicted back in the days by R.N. Gurzhi~\cite{gurzhi63,Gurzhi_1968} and realized only recently.

Drastic modification of the electron transport regime strongly affects spin, valley, and anomalous Hall effects. {Various aspects of the problem were considered in Refs.~\cite{PhysRevB.96.020401,PhysRevResearch.3.033075,PhysRevB.104.184414,PhysRevB.103.125106,PhysRevB.106.L041407,PhysRevLett.121.226601,Glazov_2021b} with the main focus on hydrodynamic regime. However, a physical picture of the effects is far from being complete. Indeed, previous works~\cite{PhysRevResearch.3.033075,PhysRevB.104.184414} were mainly concerned with the anomalous Hall hydrodynamics in three-dimensional (bulk) systems  and emphasized on spin-dependent corrections to the momentum flux tensor. The related papers in this regard are~\cite{PhysRevB.96.020401,PhysRevB.106.L041407} where the effects of rotational viscosity were studied theoretically. The paper~\cite{PhysRevB.103.125106} dealt with the two-dimensional case, but only one, the anomalous velocity (Berry curvature), contribution to the anomalous Hall effect has been considered, while the intimately related side-jump contributions were disregarded. Reference~\cite{PhysRevLett.121.226601} dealt with manifestations of the electron-electron interactions in the systems with broken Galilean or Lorentz invariance mainly focusing on the side-jump contributions. In Ref.~\cite{Glazov_2021b} a consistent theory of the spin Hall effect in ultraclean two-dimensional systems has been developed and the role of electron-electron interactions in generation and relaxation of the spin current and spin polarization was studied in detail. However, it is not \emph{a priori} clear whether the same results can be transferred to the anomalous Hall effect. A consistent theory of the anomalous Hall effect in ultraclean two-dimensional electronic channels is absent to the best of our knowledge.}

The aim of the present work is to develop a {consistent} microscopic theory of the anomalous Hall effect {that simultaneously takes into account} all relevant contributions to the Hall electric field  caused by the anomalous velocity, side-jump, and skew scattering mechanisms in ultraclean channels. We consider both ballistic regime where the mean free path exceeds by far the channels width and hydrodynamic regime where the electron-electron mean free path is much smaller than the channel width. We demonstrate that the electron scattering by impurities or phonons is a key for the anomalous Hall effect even if the electron transport at zero magnetic field is solely controlled by the electron-electron collisions and edge scattering. We show that, in contrast to the diffusive case, there is not general relation between the spin and anomalous Hall effects in ultraclean electronic channels.

The paper is organized as follows. In Sec.~\ref{sec:model} we present the general model of the anomalous Hall effect in ultraclean channels in the framework of the generalized kinetic equation. Section~\ref{sec:mechanisms} presents specific mechanisms leading to the anomalous Hall effect, and the results for the anomalous Hall field are derived in Sec.~\ref{sec:AHE} both for ballistic and hydrodynamic regimes of electron transport. Section~\ref{sec:discussion} contains the discussion of the results, relation to previous works, and routes for experimental observations of the effect. Concluding remarks are presented in Sec.~\ref{sec:concl}.

\section{\label{sec:model}Model}

\subsection{Kinetic equation}

\begin{figure}[b]
\includegraphics[width=\linewidth]{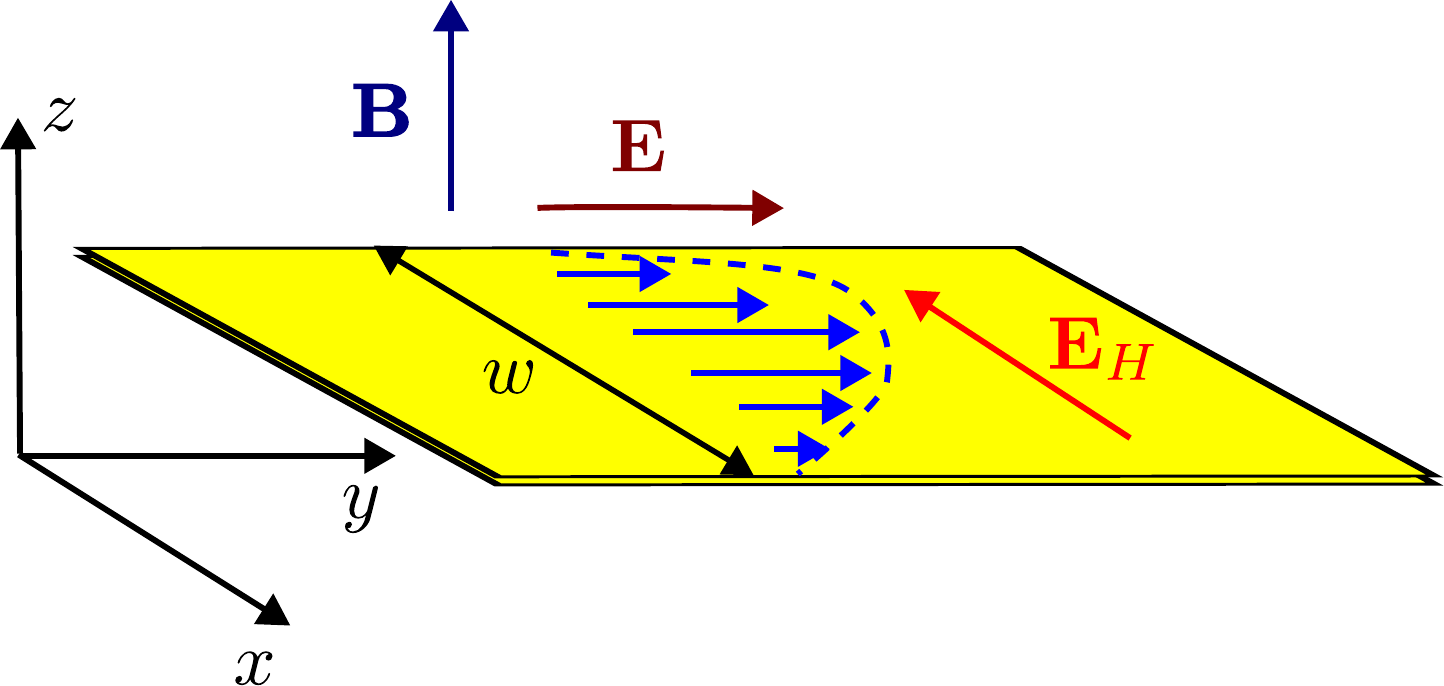}
\caption{Sketch of the studied ultraclean channel filled with a  two-dimensional electron gas. The channel width in the $x$-direction is $w$, and the channel is infinitely long along the $y$-axis. The external electric, $\mathbf E$, and magnetic, $\mathbf B$, fields are applied along the $y$ and $z$ axes, respectively. Blue arrows demonstrate the electric current profile in the hydrodynamic regime with dashed parabola illustrating its envelope, Eq.~\eqref{eq: delta f hydro}. The Hall electric field $\mathbf E_H$ is generated across the channel.}
\label{fig:channel}
\end{figure}

We consider the electron channel with width $w$ in direction of the $x$-axis and infinite length along the $y$-axis as shown in Fig.~\ref{fig:channel}. The channel is filled with the two-dimensional electron gas. We assume that the electron mean free path caused by the scattering by static impurities and phonons significantly exceeds the width of the channel $l\gg w$. %\commentMMG{I suggest to use $l$ for the scattering path as in 2D Mater.} \commentD{Ok, I've done just like that.} 
On the other hand, the channel width $w$ is much larger than the electron de Broglie wavelength that allows us to disregard formation of side-quantized subbands and consider the electron propagation semiclassically. Let us assume the diffusive scattering on the channel edges.

\begin{figure}[t]
\includegraphics[width=0.45\linewidth]{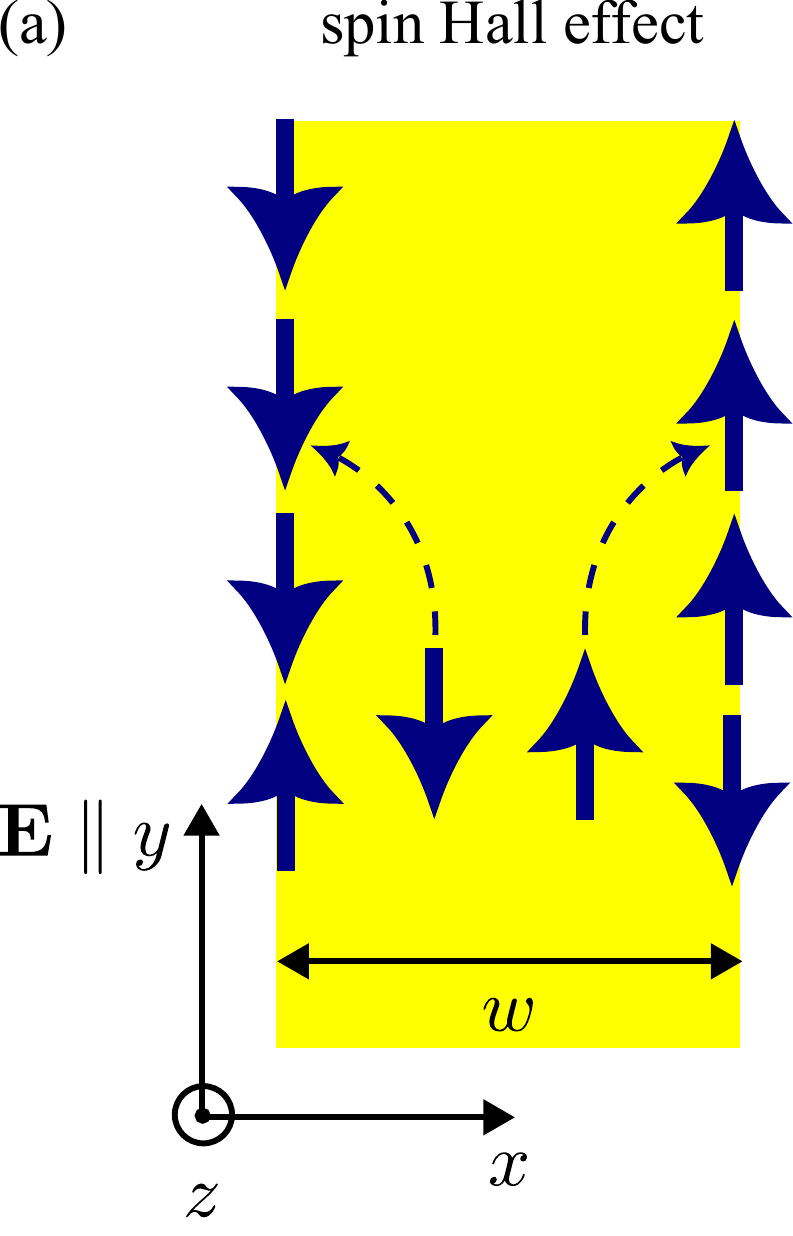}\hfill
\includegraphics[width=0.45\linewidth]{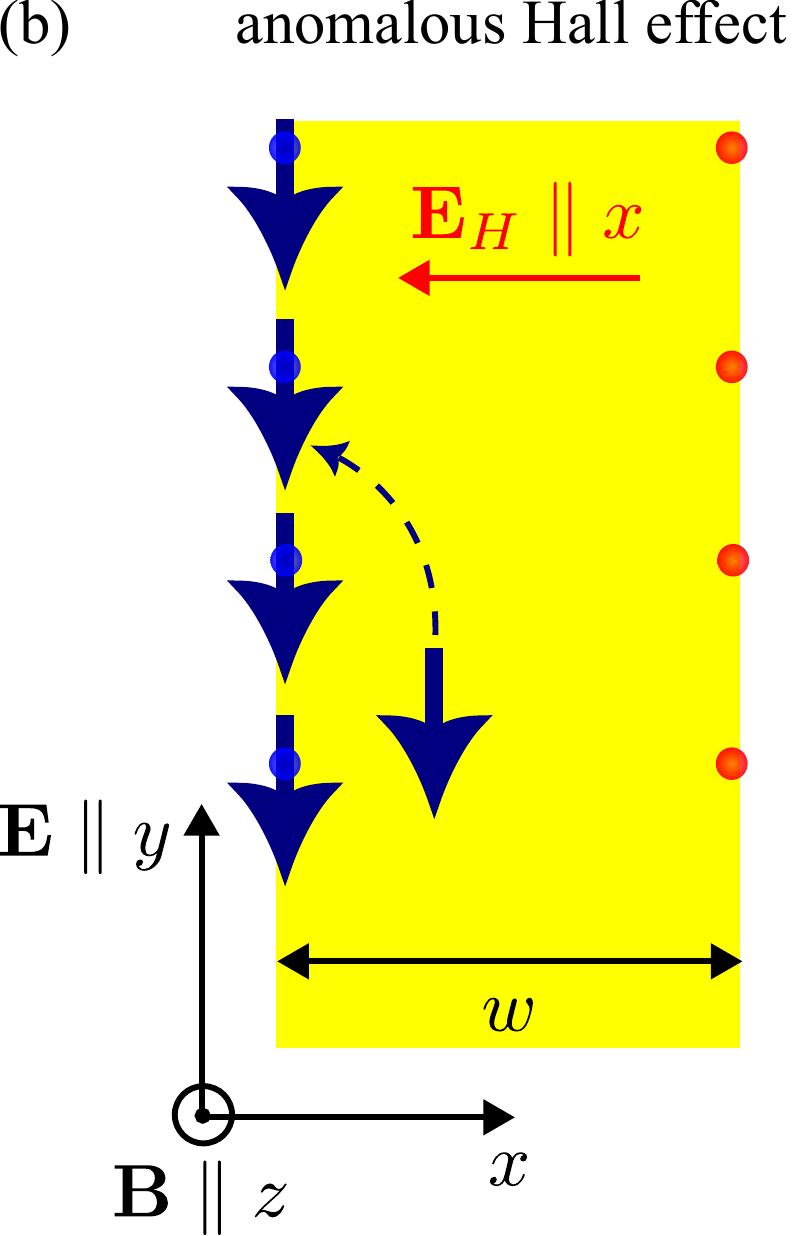}
\caption{(a) Illustration of the spin Hall effect. The spin-orbit interaction-induced anomalous velocity, side-jump and skew scattering result in the generation of the spin current in the bulk of the channel and spin polarization accumulation at the channels edges. Dark blue arrows denote spin-up ($s_z = +1/2$, uparrow) and spin-down ($s_z=-1/2$, downarrow) electrons. Due to the spin Hall effect there are more spin-up electrons at the right edge than at the left edge and vice versa, while number of electrons is the same. (b) Illustration of the anomalous Hall effect. In the presence of magnetic field $\mathbf B\parallel z$ there are more spin-down than spin-up electrons (we consider electron $g$-factor to be positive). The spin current and the spin polarization at the sample edges are converted to the electric current and charge accumulation at the sample edges. Positive and negative charges are shown by red and blue balls, respectively. This charge distribution results in the anomalous contribution to the Hall field $\mathbf E_H \parallel x$.}
\label{fig:two graphs}
\end{figure}

An external static electric field applied along the $y$-axis ($\mathbf{E} \parallel y$) results in the \emph{dc} current along the channel. Owing to the spin or valley Hall effect it results in the spin current across the channel (along the $x$-axis) and spin accumulation at the channel edges, Fig.~\ref{fig:two graphs}(a), see Refs.~\cite{dyakonov71a,kato04,wunderlich05,Ganichev2006,dyakonov_book,Glazov_2021b}. A magnetic field applied along the normal to the channel, $z$-axis ($\mathbf{B} \parallel z$), produces two effects. First, due to the Lorentz force, it generates the current along the $x$-axis, but since the electron motion is bound along this axis, the charges accumulate at the channel edges and produce an electric field which blocks the electric current. Corresponding Hall field $\mathbf E_H$ and Hall voltage $V_H$ across the channel are related to the \emph{normal Hall effect}. Second, the magnetic field owing to the Zeeman effect produces the spin polarization in the sample and converts the spin or valley currents caused by spin or valley Hall effect to the electric current. As a result, additional contributions to the charge accumulation of the channels edges and to the Hall field and voltage are formed, that is the \emph{anomalous Hall effect}, Fig.~\ref{fig:two graphs}(b).

We describe the electrons by $2\times 2$ spin density matrix $\rho_{\mathbf p} = f_{\mathbf p} \hat I + \mathbf s_{\mathbf p} \cdot \hat{\bm \sigma}$, where $\mathbf p$ is the electron momentum, $f_{\mathbf p}=\Tr\{\rho_{\mathbf p}/2\}$ is the spin-averaged electron distribution function, and $\mathbf s_{\mathbf p}$ is the spin distribution function; $\hat I$ is the unit $2\times 2$ matrix and $\hat{\bm \sigma}=(\hat \sigma_x,\hat \sigma_y,\hat\sigma_z)$ is a pseudovector composed of the Pauli matrices~\cite{dyakonov72,glazov04a}.  Under the conditions formulated above both normal and anomalous Hall effects can be found from kinetic equation for the electron spin density matrix. We present it in the form
\begin{multline}\label{eq:kin:0}
   \frac{\partial }{\partial \bold{r}}\hat{\mathbf v}_{\mathbf p} \rho_\bold{p}  +e(\bold{E}+ \mathbf E_H)\frac{\partial \rho_{\bold{p}}}{\partial \bold{p}} + \frac{e}{c}[\bold{v}\times\bold{B}]\frac{\partial \rho_{\bold{p}}}{\partial \bold{p}} \\
   =- \frac{\rho_{\bold{p}}-\overline{\rho_\bold{p}}}{\tau} + \hat{Q}_{ee}\{\rho_{\mathbf p}\} + \hat{G}_{\mathbf p}.
\end{multline}
Here $\hat{\mathbf v}_{\mathbf p}$ is the electron velocity operator that contains both kinetic and anomalous terms,  $e<0$ is the electron charge, $\tau$ is the momentum relaxation time related to the scattering by static disorder and phonons with $\overline{\rho_\bold{p}} = (2\pi)^{-1} \int_0^{2\pi} \rho_{\mathbf p} d\varphi$ being the angle-averaged density matrix and $\varphi$ being the polar angle of electron momentum. In Eq.~\eqref{eq:kin:0} $\hat{Q}_{ee}\{\rho_{\mathbf p}\}$ is the electron-electron collision integral and $\hat{G}_{\mathbf p}$ is the anomalous and spin Hall current generation rate caused by the skew scattering and side-jump effects. The form of the electron-electron collision integral $\hat{Q}_{ee}$  and generation rate $\hat{G}_{\mathbf p}$ are specified below and in Sec.~\ref{sec:mechanisms}.  Taking trace of Eq.~\eqref{eq:kin:0} we obtain the equation for the distribution function in the form
\begin{multline}\label{eq:kin. eq}
   \frac{\partial }{\partial \bold{r}}\left(\mathbf v_{\mathbf p} f_\bold{p} + \mathbf v_a s_{z,\mathbf p}\right)  +e(\bold{E}+ \mathbf E_H)\frac{\partial f_{\bold{p}}}{\partial \bold{p}} + \frac{e}{c}[\bold{v}\times\bold{B}]\frac{\partial f_{\bold{p}}}{\partial \bold{p}} \\
   =- \frac{f_{\bold{p}}-\overline{f_\bold{p}}}{\tau} +Q_{ee}\{f_{\mathbf p}\} + G_{\mathbf p},
\end{multline}
where $\bold{v}_{\mathbf p} = \bold{p}/m$  with $m$ being effective mass is the electron (kinetic) velocity, $\mathbf v_a \propto \bm    E$ is the spin-dependent anomalous velocity. General form of the kinetic equation for the spin distribution function $\mathbf s_{\mathbf p}$ can be obtained from Eq.~\eqref{eq:kin:0} by multiplying by $\hat{\bm \sigma}$ and taking the trace. It is presented for the case of $\mathbf B=0$ in Ref.~\cite{Glazov_2021b}. However, in this work we are interested in $\mathbf B$-linear effects only, thus it is sufficient to employ the kinetic equation for the $s_{z,\mathbf p}$ in the simplest form where the skew scattering and anomalous terms are disregarded
\begin{multline}
\label{eq:kin:sz}
\frac{\partial}{\partial \bold{r}}\mathbf v_{\mathbf p} s_{z,\mathbf p}+ e\bold{E}\frac{\partial s_{z,\bold{p}}}{\partial \bold{p}} \\
= - \frac{s_{z,\bold{p}}-\overline{s_{z,\bold{p}}}}{\tau} +Q_{z,ee}\{s_{z,\mathbf p}\} - \frac{s_{z,\mathbf p} - s_{z}^0}{\tau_s}.
\end{multline}
The last term in Eq.~\eqref{eq:kin:sz} describes spin relaxation towards the equilibrium spin distribution function
\begin{equation}
\label{s:z:eq}
s_{z}^0 = \frac{f^+_{p} - f^-_{p}}{2} \approx \frac{g\mu_B B}{2} \frac{df^0}{d\varepsilon}.
\end{equation}
with the phenomenological spin relaxation time $\tau_s$. Here, $f^\pm_p= f^0(\varepsilon_{p} \mp g\mu_B B/2)$ are the distribution functions of electrons with spin component being $\pm 1/2$, respectively, $f^0(\varepsilon)$ is the equilibrium Fermi-Dirac distribution, where 
\begin{equation}
\label{parabolic:disper}
\varepsilon_{p} = \frac{p^2}{2m},
\end{equation} 
is the electron dispersion, $g$ is the electron $g$-factor, and $\mu_B$ is the Bohr magneton. The last approximate equality in Eq.~\eqref{s:z:eq} is valid for small magnetic fields where the Zeeman splitting is by far smaller than the temperature.  

The kinetic equation should be supplemented with the boundary conditions. We take the simplest model of the diffusive scattering at the channel edges where~\cite{Fuchs:1938tn,sondheimer48,Falkovskii70} (see also Refs.~\cite{PhysRev.141.687,Andreev:1972wt,PhysRevB.99.035430,Glazov_2021b} for discussion of more general models):
\begin{equation}\label{eq:boundary conditions}
    \rho_{\mathbf{p}}(\pm w/2) = \begin{cases}
    \text{const},\; p_x>0,\; x=-w/2, \\
    \text{const},\; p_x<0,\; x=w/2,
    \end{cases}
\end{equation}
where $p_x=p\cos\varphi$ is the $x$-component of electron momentum.
The conditions~\eqref{eq:boundary conditions} state that the distribution of the electrons scattered from the boundary is isotropic. Additionally, we require that the electron flux through the edges is zero. 

Solution of Eqs.~\eqref{eq:kin. eq} with the boundary conditions~\eqref{eq:boundary conditions} yields non-equilibrium electron distribution function and the Hall field $\mathbf E_H$. Strictly speaking, one more equation, namely, the Poisson equation is needed to relate the imbalance of the electron density and the electric field $\mathbf E_H$. However, in realistic conditions, the contribution of the Coulomb repulsion to the electric current generation is by far stronger than that of the electron density gradient [cf. Ref.~\cite{PhysRevB.98.165440}]. Thus, we avoid solving the Poisson equation by introducing the electrochemical potential of electrons, see below.

We consider the degenerate electron gas with the Fermi energy $\varepsilon_F \gg T$ where $T$ is the temperature and we set the Boltzmann constant to be unity. Thus, kinetic phenomena are determined by the electrons in the narrow energy band $\sim T$ in the vicinity of the Fermi energy. Hence, it is convenient to use instead of the momentum and energy dependent distribution functions $f_{\mathbf p}$, $\mathbf s_{\mathbf p}$ the energy integrated functions $F_\varphi$, $\mathbf S_\varphi$ that depend on the polar angle $\varphi$ of the momentum in the form
\begin{equation}
\label{integrated}
    F_{\varphi} = \mathcal D \int_0^\infty f_{\bold{p}} d\varepsilon, \quad \mathbf S_{\varphi} = \mathcal D \int_0^\infty \mathbf s_{\bold{p}} d\varepsilon, 
\end{equation}
where $\mathcal D={m}/{2\pi\hbar^2}$ is the density of states. In the studied geometry the Hall field $\mathbf E_H\parallel x$, hence, we introduce the electric potential $\Phi(x)$ such that the Hall field 
\begin{equation}
    \label{Hall:E}
    \mathbf E_H = -\frac{\partial \Phi(x)}{\partial x} \hat{\mathbf x},
\end{equation}
with $\hat{\mathbf x}$ being the unit vector along $x$-axis and the renormalized distribution function
\begin{equation}
    \label{ren:F}
    \tilde F_\varphi(x) = F_\varphi(x) + e \mathcal D \Phi(x),
\end{equation}
where we took into account that the system is homogeneous along the $y$-axis. Formally, it is equivalent to using the electrochemical potential instead of the chemical potential for electrons.
As a result, kinetic Eq.~\eqref{eq:kin. eq} transforms to
\begin{multline}
    \label{eq:kin:1}
    \frac{\partial}{\partial x} \left(v_x \tilde F_\varphi + v_a S_{z,\varphi}^{0}\right)  + \omega_c \frac{\partial \tilde F_{\varphi}}{\partial \varphi} +\frac{\tilde F_\varphi - \overline{\tilde F_\varphi}}{\tau} + Q_{ee}\{\tilde F_\varphi\} \\
    = e \mathcal D E v_y + G_\varphi.
\end{multline}
Here we took into account that we seek the linear-in-$\mathbf E$ and linear-in-$\mathbf B$ responses only, thus we used Fermi-Dirac distribution function $f^0$ in the field term $e\mathbf E \partial f_{\mathbf p}/{\partial \mathbf p}$, so we could write down its momentum derivative as $\partial f^0/\partial \mathbf{p} = \mathbf{v}\partial f^0/\partial \varepsilon = -\mathbf{v}\delta(\varepsilon - \varepsilon_F)$;
\begin{equation}
    \label{Sz}
    S_{z,\varphi}^{0} \equiv S_z^{0} = -\frac{1}{2}\mathcal D g\mu_B B,
\end{equation}
is the equilibrium spin polarization in external magnetic field, Eq.~\eqref{s:z:eq}. It is necessary to note that in both cases where the Zeeman splitting is smaller or larger than temperature the equilibrium spin polarization \eqref{Sz} is the same provided that $|g\mu_B B| \ll \varepsilon_F$, although the approximation in \eqref{s:z:eq} is valid only in case of smaller case. The velocity components are
\begin{equation}
    \label{velocities}
    v_x = v \cos{\varphi}, \quad v_y = v\sin{\varphi},
\end{equation}
where $v= \sqrt{2\varepsilon_F/m}$ is the Fermi velocity, 
\begin{equation}
    \label{cyclotron}
    \omega_c = -\frac{eB}{mc} > 0
\end{equation}
is the cyclotron frequency, $Q_{ee}$ and $G_\varphi$ are the electron-electron collision integral and the generation rate integrated over the energy according to Eq.~\eqref{integrated}. For electron-electron collision integral acting on the particle distribution $\tilde F_\varphi$ we use the simplest form~\cite{PhysRevB.100.125419,PhysRevB.98.165412,Glazov_2021b,PhysRevB.104.085434,Narozhny:2022ud} 
\begin{equation}
    \label{Qee}
    Q_{ee}\{ \tilde F_\varphi\} = \frac{\tilde F_\varphi - \overline{\tilde F_{\varphi}}- \overline{\tilde F_{\varphi}^c}\cos{\varphi}- \overline{\tilde F_{\varphi}^s}\sin{\varphi}}{\tau_{ee}},
\end{equation}
with 
\begin{equation}
    \label{Fcs}
    \overline{\tilde F_{\varphi}^c} = \frac{1}{\pi}\int_0^{2\pi} \cos{\varphi}\tilde F_{\varphi} d\varphi, \quad    \overline{\tilde F_{\varphi}^s} = \frac{1}{\pi}\int_0^{2\pi} \sin{\varphi}\tilde F_{\varphi} d\varphi,
\end{equation}
and $\tau_{ee}$ being the electron-electron scattering time. Such simplified form of collision integral ensures the conservation of the zeroth (particle number) and first (momentum) angular harmonics of the distribution function at electron-electron collisions. Note that the description of the electron-electron scattering with a single relaxation time may not be fully appropriate~\cite{LEDWITH2019167913,PhysRevB.102.241409}, but it is sufficient for the purposes of our work.

Before turning to the anomalous Hall effect we discuss the transport regimes in the channel, provide corresponding simplifications of kinetic Eq.~\eqref{eq:kin:1} and, to illustrate the approach, reproduce known in the literature results on the normal Hall effect in the system.

Depending on the relations between the parameters $l=v\tau$, $l_{ee} = v\tau_{ee}$ and $w$ different transport regimes are realized. We are interested in two basic regimes relevant for ultraclean channels: (i) \emph{\textbf{ballistic regime}} realized at $w \ll l,l_{ee}$, and (ii) \emph{\textbf{hydrodynamic regime}}  realized at $l_{ee} \ll w\ll l$.

\subsection{Normal Hall effect}\label{subsec:normalHE}

To describe the normal Hall effect we set $v_a$ and $G_\varphi$ to be zero in Eq.~\eqref{eq:kin:1}. In the linear in the magnetic field regime we can find, first, the distribution function $\delta F_\varphi = F_\varphi - \overline{F_\varphi}$ as a first-order response to the electric field at $\omega_c=0$ and then use this function to find the first-order in $\omega_c$ contribution to $\tilde F_\varphi$.

In the \emph{ballistic regime} only the out-scattering terms in the collision integral are relevant for calculating $\mathbf E$-linear response at $\mathbf B=0$~\cite{PhysRevB.104.085434,Glazov_2021b}. Introducing the single-electron mean free path $\ell$ according to 
\begin{equation}
\label{ell}
\frac{1}{\ell} = \frac{1}{l}+\frac{1}{l_{ee}},
\end{equation}
we have at $\omega_c =0$ the electric-field induced contribution to the distribution function [cf. Refs~\cite{PhysRevB.104.085434,Glazov_2021b}]
\begin{multline}\label{eq:delta f ball}
    \delta \tilde F_{\varphi}(x) = eE\ell \mathcal D\sin\varphi\cdot \\ \times \bigg\{1-\exp\bigg[-\frac{x+w/2\;\text{sign}(\cos\varphi)}{\ell\cos\varphi}\bigg]\bigg\}.
\end{multline}
As we discuss ultraclean channels, we have $\ell \gg w$, so we can rewrite equation (\ref{eq:delta f ball}) up to first order in $w/l$:
\begin{equation}\label{eq: approx delta f ball}
        \delta \tilde F_{\varphi}(x) = eE \mathcal D\sin\varphi\frac{x+w/2\;\text{sign}(\cos\varphi)}{\cos\varphi}.
\end{equation}
Equation~\eqref{eq: approx delta f ball} is valid for $|\cos{\varphi}|\gtrsim w/\ell$, otherwise $\delta \tilde F_\varphi(x)$ rapidly drops to zero. 

The distribution function in the presence of magnetic field can be readily found substituting $\delta \tilde F_{\varphi}(x)$ in the full kinetic equation and solving it in $\omega_c$-linear regime. Following Refs.~\cite{PhysRevB.100.125419,PhysRevB.104.085434} we obtain the kinetic equation for $\Delta\tilde{F}_\varphi$ in the form
\begin{equation}\label{eq:tildeF normal ball}
    v_x\frac{\partial \Delta \tilde{F}_{\varphi}}{\partial x}+\omega_c\frac{\partial \delta \tilde F_{\varphi}}{\partial \varphi}+\frac{v}{\ell}\Delta \tilde{F}_{\varphi} = 0.
\end{equation}
and solve it taking into account diffusive boundary conditions \eqref{eq:boundary conditions} and the fact of absence of current in $x$-direction.
Ultimately, we can find Hall field taking into account that, according to Eqs.~\eqref{Hall:E} and \eqref{ren:F}, the it is is given by
\begin{equation}
\label{Hall:E:1}
E_H = -\frac{1}{e\mathcal D} \frac{\partial \overline{{\Delta}\tilde{F}_\varphi}(x)}{\partial x}.
\end{equation} 
As a result, we obtain within the logarithmic accuracy in the leading order in $\ln{\ell/w}$ 
\begin{equation}\label{Hall:normal ball}
    E_H^b = E\times \omega_c\tau\frac{w}{\pi l}\ln{\frac{\ell}{w}}.
\end{equation}
Here the subscript $b$ denotes the ballistic regime, and Eq.~\eqref{Hall:normal ball} is valid provided that $\ln{(\ell/w)} \gg 1$.  For completeness, we present the expression for the conductivity of ballistic channel~\cite{Glazov_2021b} 
\begin{equation}\label{eq:sigma_yy}
    \sigma_{yy}^b = \sigma_0 \frac{2w}{\pi l}\ln\frac{\ell}{w}, \;\;\; \sigma_0 = \frac{Ne^2\tau}{m},
\end{equation}
where $N$ is the electron density, and $\sigma_0$ is the Drude conductivity due to impurity or phonon scattering in the corresponding bulk system. {Note that logarithmic factors in Eqs.~\eqref{Hall:normal ball} and 
\eqref{eq:sigma_yy} are related to the contribution of grazing electrons: the ones with $\varphi \approx \pm \pi/2$.} These expressions agree with previous works, e.g.,~\cite{PhysRevB.100.125419,Glazov_2021b}.

In the \emph{hydrodynamic} regime where the electron-electron collisions are dominant it is sufficient to take into account only zeroth, first and second angular harmonics of the distribution function~\cite{ll6_eng,Gurzhi_1968}. We write
\begin{equation}
    \label{Fhydro}
    \delta \tilde F_{\varphi}(x) = \delta F_1 \sin{\varphi} + \delta F_2 \sin{2\varphi} \propto E,
\end{equation}
and obtain at $\mathbf B=0$ the set of equations for the coefficients $\delta F_{1,2}$:
\begin{subequations}
    \label{eq: eq.s on harmonics}
\begin{align}
        \frac{l}{2}\frac{\partial \delta F_{2}}{\partial x}+\delta F_{1}=eEl \mathcal D, \\
        \frac{l_{ee}}{2}\frac{\partial \delta F_{1}}{\partial x}+\delta F_{2}=0.
\end{align}
\end{subequations}
Solving the equation system (\ref{eq: eq.s on harmonics}) at $l\to \infty$ we obtain~\cite{PhysRevLett.117.166601}
\begin{equation}\label{eq: delta f hydro}
        \delta \tilde F_{\varphi}(x)=\frac{eE v_{y}\mathcal D}{2\eta}\left\{ \left[\left(\frac{w}{2}\right)^{2}-x^{2}\right]+\frac{2l_{ee}v_{x}}{v}x\right\},
\end{equation}
where 
\begin{equation}
    \label{visc:intro}
    \eta = vl_{ee}/4,
\end{equation} 
is viscosity of electron gas. Equation~\eqref{eq: delta f hydro} describes the Poiseuille flow of electrons in the channel. 

The magnetic field-induced correction to the distribution function can be recast as
\[
    \Delta \tilde{F}_\varphi = \Delta \tilde{F_0} +\Delta F_1\cos\varphi+\Delta F_2\cos2\varphi \propto B,
\]
and instead of Eq.~\eqref{eq:tildeF normal ball} we now have
\begin{multline}\label{eq:tildeF hydro}
     v_x\frac{\partial \Delta \tilde{F}_{\varphi}}{\partial x}+\omega_c\frac{\partial \delta F_{\varphi}}{\partial \varphi}=-\frac{\Delta F_1\cos\varphi}{\tau} -\frac{\Delta F_2\cos2\varphi}{\tau_{ee}}
\end{multline}
Solving Eq.~\eqref{eq:tildeF hydro} we obtain in agreement with Ref.~\cite{PhysRevLett.117.166601,PhysRevLett.118.226601,PhysRevB.106.245415}
\begin{equation}\label{Hall:normal hydro}
    E_H^h = E \times 2\omega_c \tau_{ee}\left\{ \frac{1}{l_{ee}^2}\left[\left(\frac{w}{2}\right)^2-x^2\right]-1\right\},
\end{equation}
where the superscript $h$ denotes the hydrodynamic regime.
Note that the conductivity in the hydrodynamic regime is given by [cf. Refs.~\cite{PhysRevLett.117.166601,Glazov_2021b}]
\begin{equation}
    \sigma_{yy}^h = \sigma_0 \frac{w^2}{12\eta \tau},
\end{equation}
and it is parametrically larger than in the ballistic regime $\sigma_{yy}^h/\sigma_{yy}^b \sim w/l_{ee}$.

\subsection{Spin distribution function}

In what follows we need the spin distribution $s_{z,\mathbf p}$ in the vicinity of the Fermi surface. Let us present the relevant equations and solutions here.
In the very low magnetic field case where 
\begin{equation}
    \label{lowBee}
    |g\mu_BB|\ll T\ll \varepsilon_F,
\end{equation} 
we can get the kinetic equation for $S_{z, \varphi}$ from Eq.~\eqref{eq:kin:sz} 
\begin{equation}
\label{eq:kin:sz:hydro+ball}
\frac{\partial}{\partial x} \cos{\varphi} S_{z,\varphi} + \frac{S_{z,\varphi} - \overline{S_{z,\varphi}}}{\ell} = e \mathcal D E \frac{S_z^0}{N} \sin{\varphi}.
\end{equation}
Importantly, the relaxation of the spin distribution function is controlled by both the electron-electron and electron-impurity and phonon collisions~\cite{glazov02,amico02,amico:045307,glazov04a,weber05,Glazov_2021b}; that is why we have $\ell = ll_{ee}/(l+l_{ee})$ in the collision integral. Noteworthy, that this equation describes both ballistic and hydrodynamic cases. Its exact solution is as follows
\begin{equation}\label{S:F:ball+hydro}
    S_{z,\varphi} = \frac{S_z^0}{N}\delta \tilde{F}_\varphi,
\end{equation}
where $\delta \tilde{F}_\varphi$ is given  by Eq.~\eqref{eq:delta f ball}.

Particularly, in the ballistic regime, $\ell \gg w$, the spin and particle distribution functions coincide (up to a numerical factor describing the spin polarization):
\begin{equation}\label{S:F:ball:b}
    S_{z,\varphi}^b = \frac{S_z^0}{N}eE \mathcal D\sin\varphi\frac{x+w/2\;\text{sign}(\cos\varphi)}{\cos\varphi},
\end{equation}
cf. Eq.~\eqref{eq: approx delta f ball}. In the hydrodynamic regime, $\ell \approx l_{ee} \ll w \ll l$ these distributions are strongly different: The particle distribution is described by the Poiselle law, Eq.~\eqref{eq: delta f hydro}, while the spin distribution has a diffusive form,
\begin{equation}
\label{Sz:hydro:B<<T}
\delta S_{z,\varphi}^h = eEl_{ee} \mathcal D \sin \varphi \frac{S_z^0}{N},
\end{equation}
which works everywhere except for the narrow stripes with the width $\sim l_{ee}$ near the edges (these stripes play no role in the following).

In the moderate magnetic field regime, where
\begin{equation}
    \label{modBee}
    T \ll  |g\mu_BB| \ll \varepsilon_F,
\end{equation} 
the  collisions between the  electrons with opposite spins are suppressed~\cite{alekseev:2} and two spin branches are basically decoupled. In that case  the kinetic equation for $S_{z,\varphi}$ takes the form
\begin{equation}
\label{eq:kin:sz:hydro+ball:T<<B}
v\cos{\varphi}\frac{\partial S_{z,\varphi}}{\partial x}   + \frac{S_{z,\varphi} - \overline{S_{z,\varphi}}}{\tau} + Q\{S_{z,\varphi}\} = e \mathcal D E v_F\frac{S_z^0}{N} \sin{\varphi},
\end{equation}
where $Q\{S_{z,\varphi}\}$ has the same for as for the particle distribution, Eq.~\eqref{Qee}. Naturally, the solution for the ballistic regime gives us the same results, Eq.~\eqref{S:F:ball+hydro}, as it was in $|g\mu_BB| \ll T$, because only the out-scattering matters. In  the hydrodynamic regime the situation is drastically different and Eq.~\eqref{eq:kin:sz:hydro+ball:T<<B} is solved similarly to Sec.~\ref{subsec:normalHE}. Its solution is given by [cf. Eq.~\eqref{eq: delta f hydro}]:
\begin{equation}\label{Sz:hydro:B>>T}
    \delta S_{z,\varphi}^{h'} = \frac{S_z^0}{N}\cdot\frac{eE \mathcal D v}{2\eta}\left\{ \left[\left(\frac{w}{2}\right)^{2}-x^{2}\right]\sin\varphi+l_{ee}x\sin2\varphi\right\}.
\end{equation}
We denote the results relevant to the hydrodynamic regime at the moderate fields condition~\eqref{modBee} with the superscript $h'$.

\section{Mechanisms of the anomalous Hall effect}\label{sec:mechanisms}

Let us now switch to the anomalous Hall effect, namely, to generation of the Hall field or voltage across the channel unrelated to the action of the Lorentz force. The key mechanisms of the anomalous Hall effect are caused by the magnetization of the sample: Largely speaking, the spin Hall current is converted, due to the magnetic field, to the electric current since the field results in the imbalance of the populations of the spin-up and spin-down electrons. As a result, the imbalance of charges arises at the channel edges and the electric field is generated,  see Fig.~\ref{fig:two graphs}. 

To calculate the anomalous Hall effect from the microscopic standpoint we need to introduce the spin-orbit interaction which mixes the charge and spin dynamics. To that end, we consider a generic two-dimensional massive Dirac model that explicitly accounts for the $\mathbf k\cdot \mathbf p$ mixing of the conduction and valence bands and includes the spin-orbit interaction~\cite{PhysRevB.75.045315,Ado_2015,2020arXiv200405091G,Glazov_2021b}. We recall that two spin branches (like in III-V or II-VI semiconductor quantum wells) or  two valleys (like in two-dimensional transition metal dichalcogenides) are described by the effective $\mathbf k\cdot \mathbf p$ Hamiltonian
\begin{equation}
    \label{H:kp}
    \mathcal H_{\pm} = \begin{bmatrix}
        0 & \pm \frac{\hbar p_{cv}}{m_0} (k_x \mp i k_y)\\
        \pm \frac{\hbar p_{cv}}{m_0} (k_x \pm i k_y) & - E_g
    \end{bmatrix}.
\end{equation}
Here the signs $\pm$ correspond to two Kramers-degenerate states (spin states with $s_z=$$\pm 1/2$ or the electron states in the time-reversal-related $\bm K_\pm$ valleys), $p_{cv}$ is the interband momentum matrix element ($p_{cv}$ is assumed to be real for simplicity), $m_0$ is the free-electron mass, and $E_g$ is the band gap. We focus on the case of small electron kinetic energies where $\varepsilon_F \ll E_g$. As a result, the electron dispersion is indeed parabolic~\eqref{parabolic:disper} with the electron effective mass at the conduction band bottom given by 
\[
m = m_0^2\frac{ E_g}{2 p^2_{cv}}.
\]      
Naturally, in the absence of magnetic field the electron dispersion is the same for the two spin branches. Importantly, this model~\eqref{H:kp} provides non-zero Berry curvature and gives rise to the anomalous transport. For $\mathbf k\to 0$ the Berry curvature reads~\cite{Sinitsyn_2007,2020arXiv200405091G}
\begin{equation}
    \label{curvature}
    \bm{\mathcal F}_\pm = \mp 2 \xi \hat{\mathbf z}, \quad \xi = \left(\frac{\hbar p_{cv}}{mE_g}\right)^2.
\end{equation}
As a result, the electrons with opposite spin projections in the presence of external electric field $\mathbf E$ acquire the \textbf{\emph{anomalous velocity}}~\cite{2020arXiv200405091G,Glazov_2021b}:
\begin{equation}\label{eq:v_a}
    \bold{v}_a^\pm = \frac{1}{\hbar}[\bm{\mathcal F}_\pm\times e\mathbf{E}] = \pm \bold{v}_a, \quad \mathbf v_a = -\frac{2\xi e}{\hbar}[\bold{\hat{z}\times E}].
\end{equation}
The anomalous velocity (\ref{eq:v_a}) does not depend on the transport regime and is the same for both ballistic and hydrodynamic regimes. The remaining contributions are related to the electron scattering processes. Let us discuss them in more detail.

We start with two `anomalous' contributions that appear due to nonzero ${\bm{\mathcal F}_\pm}$ in Eq.~\eqref{curvature} and related to the electron wavepackets shifts at the scattering. Indeed, at the elastic or quasi-elastic electron scattering by the impurity or phonon the wavepacket shifts in the real space. In the case of short-range scattering this shift is given by~\cite{belinicher82,2020arXiv200405091G,Glazov_2021b}
\begin{equation}
\label{shift}
    \bold{R_{p'p}}\equiv (X_{\bold{p'p}}, Y_{\bold{p'p}}) = \frac{\xi}{\hbar^2}\left(1+\frac{U_v}{U_c}\right)[(\mathbf p'-\mathbf p)\times\hat{\bold{z}}]
\end{equation}
Here $U_c$ and $U_v$ are the scattering matrix elements in the conduction and valence bands, respectively, $\mathbf p$ and $\mathbf p'$ are the initial and final electron momenta. In the course of electron propagation scattering events occur and the shifts given by Eq.~\eqref{shift} accumulate resulting in the \textbf{\emph{side-jump accumulation}} effect. Corresponding anomalous velocities for the charge carriers with opposite spins can be evaluated as 
\begin{multline}
    \label{int vsj:0}
    \mathbf v_{a, sj}^\pm %= \pm \mathbf v_{a,sj} \\
    =\pm\frac{2\pi}{\hbar N_\pm}\sum_{\bold{pp'}}\mathbf R_{\bold{p'p}}
    |M_\bold{p'p}|^2 \cdot \delta(\varepsilon_{p'}-\varepsilon_p) \\
    \times [f_\bold{p}(x) \pm s_{z,\bold{p}}(x)],
\end{multline}
where $M_{\bold{p'p}}$ is scattering matrix element and $N_\pm = \mathcal D \varepsilon_F \pm S_{z}$ are the occupancies of the spin branches.

As a result, the side-jump contribution to the anomalous velocity in Eqs.~\eqref{eq:kin. eq} and \eqref{eq:kin:1} read
\begin{multline}
    \label{int vsj}
    \mathbf v_{a, sj}
    =\frac{4\pi}{\hbar S_z}\sum_{\bold{pp'}}\mathbf R_{\bold{p'p}}
    |M_\bold{p'p}|^2 \cdot \delta(\varepsilon_{p'}-\varepsilon_p) %\\
     s_{z,\bold{p}}(x).
\end{multline}
Next, we evaluate by virture of Eq.~\eqref{int vsj} the side-jump accumulation velocity in the ballistic and hydrodynamic regimes. In the ballistic regime, using approximation \eqref{S:F:ball:b} we obtain in the leading order in $w/l$:
\begin{equation}\label{v_sj ball}
    \mathbf v_{a,sj}^b =\left(1+\frac{U_v}{U_c}\right)\frac{w}{\pi l}\ln \left(\frac{\ell}{w}\right)\frac{2\xi }{\hbar}[\hat{\mathbf z}\times e\mathbf E].
\end{equation}
The logarithm, similarly to the case of conductivity, Eq.~\eqref{eq:sigma_yy}, results from the grazing electrons contribution and the pre-factor $\propto l^{-1}$ due to the fact that only impurity and phonon scattering contribute to the side-jump velocity: The electron-electron interaction in the systems with parabolic dispersion cannot result in the electric current generation or relaxation.
In the hydrodynamic regime at small magnetic fields, $|g\mu_B B| \ll T$, Eq.~\eqref{lowBee}, to find the side-jump accumulation velocity we use the approximation \eqref{Sz:hydro:B<<T} with a result
\begin{subequations}
    \label{vsj:hydro:tot}
 \begin{equation}
\label{v_sj hydro B<<T}
\mathbf v_{a,sj}^{h} = \left(1+\frac{U_v}{U_c}\right)\frac{l_{ee}}{l}\frac{\xi}{\hbar}[\hat{\mathbf z}\times e\mathbf E].
\end{equation}
Here, again, a prefactor contains $l^{-1}$ because only impurity or phonon scattering contributes to the side-jump accumulation.
In the moderate field case, Eq.~\eqref{modBee}, using Eq.~\eqref{Sz:hydro:B>>T}, we obtain
\begin{equation} \label{v_sj hydro B>>T}
    \mathbf v_{a,sj}^{h'} = \left(1+\frac{U_v}{U_c}\right)\frac{2}{ll_{ee}}\frac{\xi}{\hbar}\left[\left(\frac{w}{2}\right)^2-x^2\right][\hat{\mathbf z}\times e\mathbf E].
\end{equation}
\end{subequations}

Shifts of the wavepackets at the scattering are also accompanied by a change in the electron energy. As a result, the scattering by impurities becomes asymmetric and can lead to the current generation. This \emph{\textbf{anomalous distribution}} contribution results in the generation rate in the form (we recall that $\mathbf E \parallel \hat{\mathbf y}$)~\cite{2020arXiv200405091G,Glazov_2021b}
\begin{multline}\label{Gadistpm int. form}
    G_{\varphi,adist} = \frac{2\pi}{\hbar} \mathcal D\int_0^\infty d\varepsilon_p \times\sum_{\bold{p'}}|M_{\mathbf p'\mathbf p}|^2 \\ \cdot(eEY_{\bold{p'p}}) \cdot\delta'(\varepsilon_{p'}-\varepsilon_p)[s_{z,\mathbf p'}^{0}-s_{z,\mathbf p}^{0}].
\end{multline}
As it can be seen from Eq.~(\ref{Gadistpm int. form}), the anomalous distribution contribution arises due to scattering by impurities and is calculated using the equilibrium spin distribution function. Hence, it does not depend on transport regime, so its generation rate is the same for both ballistic and hydrodynamic regimes:
\begin{equation}\label{Gadistpm}
    G_{\varphi,adist} = -\bigg(1+\frac{U_v}{U_c}\bigg)\frac{2\xi e}{l\hbar}E S_z^{0}\cos\varphi.
\end{equation}

Finally, the asymmetric or \emph{\textbf{skew scattering}} effect results in the spin Hall current and, in the presence of magnetic field, in the anomalous Hall effect. Similarly to the side-jump effect, the skew scattering takes place at the impurity or phonon scattering only. Following Refs.~\cite{2020arXiv200405091G,Glazov_2021b} we obtain
\begin{multline}
    G_{\varphi, skew} =  \xi S_{imp}\mathcal D \\  \times{\frac{1}{\hbar^2}}\int_0^\infty d\varepsilon_{p} \sum_\bold{p'}[\bold{p\times p'}]_z \delta(\varepsilon_p-\varepsilon_p')\delta s_{{z},\bold{p'}}(x),
\end{multline} 
with~\footnote{{Here we correct a misprint in the expression for $S_{imp}$ in Ref.~\cite{Glazov_2021b}}.}
\begin{equation}\label{eq:S_imp}
    S_{imp} = \frac{2\pi U_v}{\tau}+4\frac{U_v}{U_c}\frac{\hbar}{\mathcal D\varepsilon_F \tau^2}.
\end{equation}
Here first term in the right side of (\ref{eq:S_imp}) is responsible for scattering in the third order of impurity potential, and second term is responsible for two-impurity coherent scattering~\cite{Ado_2015} or two-phonon scattering~\cite{2020arXiv200405091G}. Making use of Eqs.~\eqref{S:F:ball+hydro} and \eqref{eq: approx delta f ball} for the ballistic regime we obtain 
\begin{equation}\label{Gskpm ball}
    G_{\varphi, skew}^{b} =  S_{imp}\frac{\tau^2}{\hbar}\frac{w}{\pi l}\ln\left(\frac{\ell}{w}\right)\frac{\xi e}{l\hbar}EN S_z^{0} \cos\varphi.
\end{equation}
Similarly to the case of side-jump accumulation the logarithmic factor accounts for the grazing electrons and, as before, we took into account the leading in the $w/l \ll 1$ contribution with a big factor $\ln(\ell/w) \gg 1$.
In the hydrodynamic transport regime we use the spin distribution function either in the form of Eq.~\eqref{Sz:hydro:B<<T} [weak fields, Eq.~\eqref{lowBee}] or Eq.~\eqref{Sz:hydro:B>>T} [moderate fields, Eq.~\eqref{modBee}] to obtain
\begin{subequations}
    \label{Gskpm:hydro:tot}
\begin{equation}\label{Gskpm hydro B<<T}
    G_{\varphi,skew}^{h} =  S_{imp}\frac{\tau^2}{\hbar}\frac{l_{ee}}{l}\frac{\xi e}{l\hbar}EN_{0} S_z^{0} \cos\varphi,
\end{equation}
and
\begin{multline}\label{Gskpm hydro B>>T}
    G_{\varphi,skew}^{h'} =  S_{imp}\frac{\tau^2}{\hbar}\frac{\xi e}{\hbar l} EN S_z^{0} \\
    \times \frac{1}{ll_{ee}}\left[\left(\frac{w}{2}\right)^2-x^2\right] \cos\varphi,
\end{multline}
respectively.
\end{subequations}

It is worth mentioning that in the presence of magnetic field the $\mathbf B$-linear contributions to the electron-impurity or electron-phonon scattering matrix elements appear in the form of $\propto [\mathbf p \times \mathbf p'] \cdot \mathbf B$, cf. Refs.~\cite{PhysRevB.77.085328,Kibis:1999aa} where $\mathbf k \mathbf B$-linear terms were discussed. These contributions unrelated to the spin polarization contain a smallness $\sim \hbar\omega_c/E_g$, and we disregard such contributions.

\section{\label{sec:AHE} Anomalous Hall effect}

Now we turn to the calculation of the anomalous Hall field and voltage from Eq.~\eqref{eq:kin:1} which can be recast in the form:
\begin{multline}
    \label{eq:kin:1:AHE}
    \frac{\partial}{\partial x} \left(v_x \tilde F_\varphi + v_a S_{z,\varphi}^{0}\right)  %+ \omega_c \frac{\partial \tilde F_{\varphi}}{\partial \varphi} 
    +\frac{\tilde F_\varphi - \overline{\tilde F_\varphi}}{\tau} + Q_{ee}\{\tilde F_\varphi\} %\\
    = %e \mathcal D E v_y + 
    G_\varphi,
\end{multline}
where we omitted the terms $\propto E$ and $\propto \omega_c$ that describe the normal Hall effect studied in Sec.~\ref{subsec:normalHE}. Both the anomalous velocity contribution in the left hand side, $\propto v_a S_{z,\varphi}$ and the generation contribution in the right hand side, $G_\varphi$, are proportional to the spin polarization giving rise to the anomalous Hall effect. Below we solve Eq.~\eqref{eq:kin:1:AHE} to the lowest order in $S_z$ in the ballistic and hydrodynamic regimes, determine $\overline{\tilde F_\varphi}$ from the solution of kinetic equation and calculate the anomalous Hall field by means of Eq.~\eqref{Hall:E:1}.

\subsection{\label{sec:AHE ball}Anomalous Hall effect in ballistic regime}
The kinetic Eq.~\eqref{eq:kin:1:AHE} in the ballistic regime can be transformed as
\begin{equation}
    \label{eq:kin:1:AHE:ball}
        \frac{\partial}{\partial x} \left(v_x \tilde F_\varphi + v_a S_{z,\varphi}^{0}\right)   + \frac{v}{l}(\tilde F_{\varphi} - \overline{\tilde F_\varphi}) =  
    G_\varphi^{b}.
\end{equation}
We recall that $v_a$ contains both contributions from Eqs.~\eqref{eq:v_a} and~\eqref{v_sj ball}. Note that in the calculation of the Hall field we cannot disregard the out-scattering terms in the collision integral. However, in our situation we can neglect the electron-electron collisions since, as calculation shows, only zeroth and first angular harmonics of $\tilde F_{\varphi}$ are now relevant, cf. Eq.~\eqref{eq:tildeF normal ball}.
This equation can be solved using the method developed in Ref.~\cite{Glazov_2021b}. 

First, we calculate anomalous contributions caused by both the anomalous velocity and side-jump accumulation. Using the boundary conditions of the electric current absence through the edges we get
\begin{equation}
\label{F1}
    F_1 = -\frac{2v_a}{v} S_{z,\varphi}^{0},
\end{equation}
where $F_1$ is the first angular harmonic of $\tilde F_\varphi$. Equation~\eqref{F1} means that there is no electric current along the $x$-axis. The complete solution reads 
\begin{equation}
     \tilde{F}_\varphi^{a} = 2\frac{v_{a}x}{vl}S_z^0 - 2\cos\varphi \frac{v_{a}}{v}S_z^0.
\end{equation}
As discussed above, in the ballistic regime, anomalous velocities do not depend on the coordinate, so the contributions to the Hall field from anomalous velocities take the simple form:
\begin{equation}\label{AHE:a/sj ball}
    E_{H,a}^{b} = -\frac{2S_z^0}{e\mathcal D}\frac{v_{a}^b}{vl}.
\end{equation}

Second, we calculate the contribution stemming from the generation term. It follows from Eqs.~\eqref{Gadistpm} and \eqref{Gskpm ball} that both the anomalous distribution and skew scattering-related generation rates have the same form, namely,
\[
G^b_\varphi =  (G^b_{adist}+G^b_{skew})\cos{\varphi},
\]
with angular-independent prefactors $G_{adist/skew}$, Eqs.~\eqref{Gadistpm} and \eqref{Gskpm ball}.
Correspondingly, in the leading order in $w/\ell$ we get
\begin{equation}
    \tilde{F}_\varphi^{adist/skew} = \frac{x}{v} G^b_{adist/skew},
\end{equation}
where, as seen from Eqs.~\eqref{Gadistpm} and \eqref{Gskpm ball}, the generation rates of $G_{adist/skew}$ also do not depend on the coordinates. Accordingly, their contributions to the Hall field have the form:
\begin{equation}\label{AHE:sk/adist ball}
    E_{H,adist/skew}^b = -\frac{1}{ev\mathcal D}G^b_{adist/skew}.
\end{equation}
Equations~\eqref{AHE:a/sj ball} and \eqref{AHE:sk/adist ball} are solutions to our problem in ballistic regime.

\subsection{\label{sec:AHE hydro}AHE in hydrodynamic regime}

Now let us solve Eq. \eqref{eq:kin:1:AHE} in the hydrodynamic regime. Similarly to the case of the normal Hall effect studied in Sec.~\ref{subsec:normalHE}, we keep, in the distribution function, the angular harmonics up to second: 
\begin{equation}\label{expension hydro}
    \tilde{F}_\varphi = \overline{\tilde{F}_\varphi}+F_1\cos\varphi + F_2\cos2\varphi. 
\end{equation}
Analogously to derivation of Eqs.~\eqref{eq:tildeF hydro} and \eqref{eq:tildeF hydro} we transform Eq.~\eqref{eq:kin:1:AHE} as
\begin{multline}\label{eq:kin:1:AHE:hydro}
    \frac{\partial}{\partial x}\Big(v_x \tilde{F}_\varphi+v_aS_{z,\varphi}\Big) =\\ -\frac{F_1\cos\varphi}{\tau} - \frac{F_2\cos2\varphi}{\tau_{ee}}  + G_\varphi^{h}.
\end{multline}
Here $v_a$ is the sum of anomalous velocities \eqref{eq:v_a} and \eqref{vsj:hydro:tot}, and the generation rate $G_\varphi^h$ is given similarly to the ballistic case in a form
\[
G^h_\varphi =  (G^h_{adist}+G^h_{skew})\cos{\varphi} \equiv G_h\cos\varphi,
\]
with angle-independent $G_{adist/skew}^h$. Combining Eqs.~\eqref{expension hydro} and \eqref{eq:kin:1:AHE:hydro} we obtain a set of three coupled equations:
\begin{subequations}
\label{hydro:AHE:set:set}
    \begin{equation}\label{eq. system anomalous hydro 0}
        \frac{\partial}{\partial x}\left(\frac{v}{2} F_1 + v_a S_z^0\right) = 0,
    \end{equation}
    \begin{equation}\label{eq. system anomalous hydro 1}
        \frac{\partial}{\partial x}\left[v\left(\overline{\tilde{F}_\varphi}+\frac{F_2}{2}\right)\right] = -\frac{F_1}{\tau} + G_h,
    \end{equation}
    \begin{equation}\label{eq. system anomalous hydro 2}
        \frac{\partial}{\partial x}\left(\frac{v}{2}F_1\right) + \frac{F_2}{\tau_{ee}}=0.
    \end{equation}
\end{subequations}
Using boundary conditions of vanishing current through the channel edges we get the first harmonic from \eqref{eq. system anomalous hydro 0}:
\begin{equation}\label{F1: anomalous}
   F_1 = - \frac{2v_a}{v} S_z^0,
\end{equation}
cf. Eq.~\eqref{F1}.
Using this first harmonic we can get second harmonic from equation \eqref{eq. system anomalous hydro 2}:
\begin{equation}\label{F2: anomalous}
    F_2 = \tau_{ee}\frac{\partial}{\partial x}\left(v_a S_z^0\right).
\end{equation}
Substituting this expression to Eq.~\eqref{eq. system anomalous hydro 1} we have
\begin{equation}\label{F0: anomalous}
    v\frac{\partial \overline{\tilde{F}_\varphi}}{\partial x} = \frac{2v_a S_z^0}{l} + G_h - \frac{l_{ee}}{2}\frac{\partial^2}{\partial x^2}(v_a S_z^0).
\end{equation}
Solving this equation we can find $\overline{\tilde{F}_\varphi}$ and the Hall field in the form
\begin{multline}\label{AHE:a hydro}
    E_{H,a}^h = -\frac{2S_z^0}{e\mathcal D}\frac{v_a^h}{vl}+\frac{\tau_{ee}}{2e\mathcal{D}}\frac{\partial^2}{\partial x^2}(v_a S_z^0)\\
    -\frac{1}{ev\mathcal D}\left(G_{skew}^h + G_{adist}^h\right).
\end{multline}
Equation \eqref{AHE:a hydro} presents the solution to our problem in hydrodynamic regime. Note that in the case of weak magnetic fields, Eq.~\eqref{lowBee}, the anomalous velocity is coordinate independent and $F_2$ vanishes.

\section{Discussion}\label{sec:discussion}

Equations~\eqref{AHE:a/sj ball}, \eqref{AHE:sk/adist ball}, and \eqref{AHE:a hydro} are the main results of this work. They present the anomalous Hall field in ballistic and hydrodynamic transport regimes caused by the anomalous velocity, side-jump accumulation and anomalous distribution, and skew scattering mechanisms. Here we briefly analyze and discuss the obtained results.

\begin{figure*}[t]
    \centering
    \includegraphics[width=0.85\textwidth]{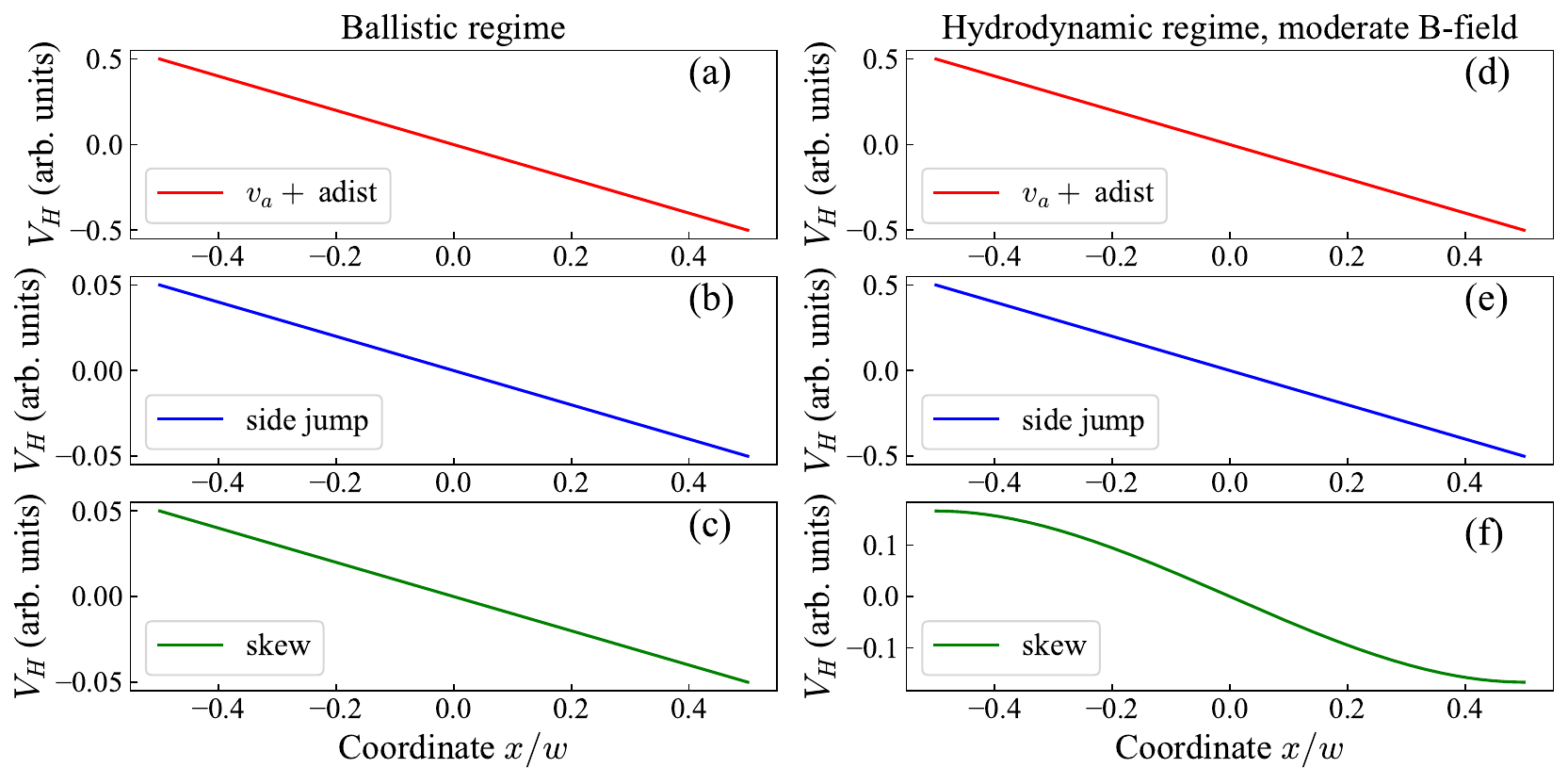}
    \caption{Anomalous Hall voltage $V_H$ distribution across the sample calculated for the ballistic regime (left panels a-c) and for hydrodynamic regime at moderate magnetic fields, Eq.~\eqref{modBee} (right panels d-f). Individual contributions from anomalous velocity and anomalous distribution (a,d), side jump accumulation (b,e), and skew scattering (c,f) are shown. Plots are given not to scale.}
    \label{fig:VH}
\end{figure*}

\subsection{Ballistic regime}

Let us recall the main contributions to the anomalous Hall effect for the ballistic regime derived in Sec.~\ref{sec:AHE ball}. Since there is a small parameter
 $(w/l)\ln (\ell/w) \ll 1$ in the ballistic transport regime, the main contributions to the anomalous Hall field are given by the anomalous velocity and anomalous distribution, Eqs.~\eqref{AHE:a/sj ball} and \eqref{AHE:sk/adist ball}, and can be recast as 
\begin{subequations}
\label{AHE:ball:res}
\begin{equation}\label{Hall: amain ball}
   E_{H}^b = E_{H,a}^b+E_{H,adist}^b = -\frac{2\xi}{\mathcal D v\hbar}\left(1-\frac{U_v}{U_c}\right)ES_z^0\frac{1}{l}.
\end{equation}  
The remaining contributions from asymmetric scattering and side jump contain a small parameter $w/l\ln(\ell/w)\ll 1$, so they are important only if $U_c \approx U_v$ where $E_H^b$ in Eq.~\eqref{Hall: amain ball} becomes negligible. These sub-leading terms read:
\begin{multline}\label{E_H sj+sk}
   {E_H^{b'}= E_{H,sj}+E_{H, skew} = \frac{2\xi}{\mathcal Dv\hbar}\frac{w}{\pi l}\ln\left(\frac{\ell}{w}\right) }\\ {\times \left\{2\left(1 - \frac{U_v}{U_c}\right) -\frac{\pi U_v\tau N}{\hbar}\right\}ES_z^0\frac{1}{l}.}
\end{multline}
\end{subequations}
Similar conclusions were drawn in the Ref.~\cite{Glazov_2021b} regarding the relationships between the contributions to the spin/valley current in the absence of a magnetic field. It is obvious that the scattering is suppressed in an ultraclean channels, so the contribution from the anomalous velocity should be the leading one. {Among the scattering-related contributions, the dominant one is due to the anomalous distribution: the generation rate of the anomalous distribution [Eqs.~\eqref{Gadistpm int. form} and \eqref{Gadistpm}] is proportional to the scattering rate in the bulk, $\propto 1/l$, and corresponding contribution to the anomalous Hall field $\propto 1/l$ just like that of the anomalous velocity, Eq.~\eqref{Hall: amain ball}. Note that this contribution results from the equilibrium distribution function with account for the electric field in the energy conservation $\delta$-function in the collision integral~\eqref{Gadistpm int. form}. By contrast, the side-jump accumulation and skew scattering have both the small factor $\propto 1/l$ due to scattering in the bulk stemming from the generation rate and, additionally, the smallness $(w/l)\ln (\ell/w) \ll 1$ resulting from the electric field induced $S_{z,\varphi}$.}

{Similarly to diffusive systems, at $U_v = U_c$ the only contribution of skew scattering in the third order of the impurity potential remains. In this case the side-jump anomalous velocity and two-impurity coherent scattering contributions of skew scattering compensate each other similarly to anomalous velocity and anomalous distribution \cite{Glazov2020b}. We also note that,} in contrast to systems with diffusive propagation of electrons, in ballistic channels, as follows from Eq.~\eqref{Hall: amain ball}, there is no complete compensation of the anomalous velocity contribution. This is due to the fact that a noticeable part of the momentum is lost on the channel edges, and not in its bulk, while the anomalous velocity is present everywhere, see~\cite{Glazov_2021b} for details.

Interestingly, the normal contribution to the Hall field \eqref{Hall:normal ball} has a small parameter $(w/l)\ln (\ell/w) \ll 1$ in the ballistic regime unlike the anomalous velocity and anomalous distribution contributions. However, the latter ones have a small parameter $\propto \xi \propto 1/E_g^2$. In the case when the scattering potentials in the conduction and valence band are the same, $U_c = U_v$, the anomalous Hall effect is further suppressed and has both small parameters: $(w/l)\ln (\ell/w) \ll 1$ and  $\xi$.

For all mechanisms of the anomalous Hall effect the Hall voltage is a linear function of the coordinate $x$ in the ballistic regime as it is illustrated in Fig.~\ref{fig:VH}(a-c) where the dependence $V_H(x)$ is plotted for individual mechanisms by taking integrals of Eqs.~\eqref{AHE:ball:res} over the coordinate.
It is noteworthy that in ballistic channels Eqs.~\eqref{AHE:ball:res} and corresponding expressions for the distribution of the Hall voltage across the channel can be derived directly from Eqs. (36), (37), (39), (40), and (42) of Ref.~\cite{Glazov_2021b}. In fact, in such a case the spin polarization obeys the same equation as $\tilde F_\varphi(x)$. Introducing the equilibrium magneto-induced electron spin polarization as 
\begin{equation}
\label{Ps}
P_s = 2\frac{S_z^0}{N},
\end{equation}
we obtain, up to    a constant, the Hall voltage
\begin{equation}
\label{V_H:b}
V_H^b(x) = \frac{1}{e \mathcal D} P_s  \Delta N^+_{b}(x),
\end{equation}
where $\Delta N^{+}_{b} = - \Delta N^-_b$ are the densities of electrons with spin $\pm 1/2$ components in the absence of magnetic field generated due to the spin or valley Hall effect and introduced in Ref.~\cite{Glazov_2021b}. As we see below this universality does not hold in the hydrodynamic regime. It also breaks in deeply diffusive systems where spin relaxation controls spin polarization distribution in the spin Hall effect~\cite{dyakonov71a,dyakonov_book} and is virtually unimportant for the anomalous Hall effect.

\subsection{Hydrodynamic regime}

Let us present the complete expressions for the anomalous Hall fields in the hydrodynamic transport regime substituting microscopic expressions for the velocities and generation rates into Eq.~\eqref{AHE:a hydro}. Contributions from the anomalous velocity and anomalous distribution are the same as in the ballistic regime studied above, see Eq.~\eqref{Hall: amain ball}. Naturally, they have the same form both for the low \eqref{lowBee} and moderate \eqref{modBee} magnetic fields:
\begin{subequations}
    \label{AHE:hydro:res}
\begin{equation}\label{Hall: va + adist hydro}
    E_{H,a}^{h/h'}+E_{H,adist}^{h/h'} = -\frac{2\xi}{\mathcal Dv\hbar l}\left(1-\frac{U_v}{U_c}\right)ES_z^0.
\end{equation}
These contributions do not have coordinate dependence. Other contributions are coordinate independent only in case of low magnetic field. The side jump contribution to the Hall field reads
\begin{equation}
\label{Hall: sj hydro':0}
    E_{H,sj}^{h} = \frac{2\xi}{\mathcal Dv\hbar l}\frac{l_{ee}}{l} \left(1+\frac{U_v}{U_c}\right)ES_z^0,
\end{equation}
at low magnetic fields, Eq.~\eqref{lowBee}, and
\begin{equation}
\label{Hall: sj hydro'}
    {E_{H,sj}^{h'} = \frac{2\xi}{\mathcal Dv\hbar l}\left(1+\frac{U_v}{U_c}\right)ES_z^0,}
\end{equation}
at moderate magnetic fields, Eq.~\eqref{modBee}, respectively. {The side-jump contribution to the anomalous Hall field is coordinate independent both in weak and moderate magnetic fields. The $E_{H,sj}^{h'} / E_{H,sj}^h = l/l_{ee} \gg 1$ because electron-electron collisions between the particles with opposite spins, if active (weak fields), suppress spin current.} Expressions for the skew scattering contribution to the Hall field are, respectively,
\begin{equation}
\label{pre:last:hydro}
    E_{H,sk}^h = -\frac{\xi}{\mathcal Dv \hbar l} S_{imp}\frac{\tau^2}{\hbar}\frac{l_{ee}}{l}EN_{0} S_z^{0},
\end{equation}
for low magnetic fields, Eq.~ \eqref{lowBee}, and 
\begin{multline}
\label{last:hydro}
    E_{H,sk}^{h'} = -\frac{\xi}{\mathcal Dv \hbar l} S_{imp}\frac{\tau^2}{\hbar} EN S_z^{0} \\
    \times \frac{1}{ll_{ee}}\left[\left(\frac{w}{2}\right)^2-x^2\right],
\end{multline}
\end{subequations}
for moderate fields, Eq.~\eqref{modBee}. {Making use of Eqs.~\eqref{Hall: va + adist hydro} -- \eqref{last:hydro} we obtain the Hall voltage in the form
\begin{subequations}
\begin{equation}\label{V:H:hydro low}
    V_H^h(x) = -\frac{2\xi}{\hbar l}\frac{ES_z^0}{v\mathcal D}\left(\frac{U_v}{U_c} - 1 - \frac{\pi U_vN\tau_{ee}}{2\hbar}\right)\cdot x
\end{equation}
in case of low magnetic fields \eqref{lowBee} and
\begin{multline}\label{V:H:hydro mod}
    V_H^{h'} = -\frac{4\xi}{\hbar l}\frac{ES_z^0}{v\mathcal D}\\ \times\left\{\frac{U_v}{U_c} - \frac{\pi U_vN\tau_{ee}}{2\hbar}\cdot \frac{1}{l_{ee}^2}\left[\left(\frac{w}{2}\right)^2 - \frac{x^2}{3}\right]\right\}\cdot x
\end{multline}
\end{subequations}
in case of moderate magnetic fields \eqref{modBee}.  In case of low magnetic fields \eqref{V:H:hydro low} and $U_v = U_c$, only the single impurity third order skew scattering contribution remains as we discussed previously in for the ballistic regime \eqref{AHE:ball:res}.}
Figure~\ref{fig:VH}(d-f) shows the Hall voltage distribution across the channel for the hydrodynamic regime and moderate magnetic fields (in the weak field regime $V_H(x) \propto x$ as in the ballistic regime). In our calculation we assumed that $ll_{ee} \ll w^{2}$ (cf. Ref.~\cite{Glazov_2021b}), in this case the skew scattering contribution has a nonlinear coordinate dependence.

Similarly to the ballistic regime, in the hydrodynamic regime the anomalous Hall field has a smallness $\propto (\varepsilon_F/E_g)^2$ since it requires the spin-orbit interaction. In case of low magnetic field the side-jump and skew scattering contributions also have smallness $\propto l_{ee}/l$, for the moderate fields where $|g\mu_B B| \gg T$ this smallness is absent.

Interestingly, in the hydrodynamic regime the Hall field appears due to the impurity or phonon scattering and it vanishes in the purely hydrodynamic case where $l\to \infty$ as it follows from Eqs.~\eqref{AHE:hydro:res}. Physically, it is because the electron electron collisions conserve the total momentum of the colliding pair and cannot, for the parabolic dispersion, result in the electric current relaxation or generation due to effective Gallilean invariance of the system. Similarly, for the parabolic dispersion the side-jump at the electron-electron collisions is inefficient~\cite{PhysRevLett.121.226601}. It is in stark contrast with the spin current generation and spin accumulation at the channel edges where the electron-electron collisions play a crucial role~\cite{Glazov_2021b,PhysRevB.106.235305}.

{It is instructive to analyze the obtained results in the hydrodynamic regime from a more general viewpoint and establish its relations with previous works~\cite{PhysRevB.96.020401,PhysRevResearch.3.033075,PhysRevB.104.184414,PhysRevB.103.125106,PhysRevB.106.L041407,PhysRevLett.121.226601} on `anomalous' hydrodynamics. To that end, we allow slow (on the time scale of $\tau_{ee}$) and smooth (on the length scale $l_{ee}$) variations of electron density $N(\mathbf r,t)$ and hydrodynamic velocity $\mathbf u(\mathbf r, t) = N^{-1} \sum_{\mathbf p} (\mathbf p/m) f_p$. 
Combining equations~\eqref{eq: eq.s on harmonics} and \eqref{hydro:AHE:set:set} we arrive at the 
standard set of hydrodynamic equations for a fluid~\cite{ll6_eng}, namely, of continuity equation\footnote{{In the reasonable limit of incompressible flow Eq.~\eqref{cont:hydro} requires $\bm \nabla \cdot (\mathbf u+ \mathbf{u}_a)$. The anomalous velocity should be taken into account in the boundary conditions (while in the bulk of the channel its divergence vanishes, cf. Ref.~\cite{PhysRevB.103.125106}).}}
\begin{subequations}
    \label{set:hydro:anomalous}
    \begin{equation}
    \label{cont:hydro}
    \frac{\partial N}{\partial t} + \bm \nabla \cdot N (\mathbf u+ \mathbf{u}_a) =0,
    \end{equation}
and linearized equation for the hydrodynamic velocity 
    \begin{multline}
    \label{Euler:hydro}
    \frac{\partial N\mathbf{u}}{\partial t} + N{\mathbf u \times \bm \omega_c}  \\
    = -\bm \nabla \hat{\Pi} - \frac{N\mathbf{u}}{\tau} + \frac{e N}{m}(\mathbf E + \mathbf{E}_H) + \mathbf f.     
    \end{multline}
\end{subequations}
Here, $\mathbf u_a$ is the anomalous velocity (related both to Berry curvature and side-jump accumulation), $\hat{\Pi}$ is the momentum flux density tensor, and the force density $\mathbf f$ results from the anomalous distribution and skew scattering induced current generation rates. Equation~\eqref{eq. system anomalous hydro 2} provides the expression for the components of $\hat\Pi$ that, in our two-dimensional case, reduces to 
\begin{equation}
    \label{Pi:standard}
    \Pi_{\alpha\beta} = -N \eta_{\alpha\beta} \left(\frac{\partial u_\alpha}{\partial r_\beta} +\frac{\partial u_\beta}{\partial r_{\alpha}} -  \delta_{\alpha\beta} \frac{\partial u_\gamma}{\partial r_\gamma}\right),
\end{equation}
with the viscosity coefficients determined by the electron-electron collisions
\begin{equation}
    \label{eta:ee}
\eta_{xx} = \eta_{yy} \equiv \eta = \frac{v l_{ee}}{4}, \quad \eta_{xy},~\eta_{yx}=0.
\end{equation}
in agreement with Eq.~\eqref{visc:intro}.  One can readily check, that solution of Eqs.~\eqref{set:hydro:anomalous} with no-slip boundary conditions $u_y(\pm w/2) =0$ and appropriate expressions for $\mathbf u_a = v_a S_z^0 \hat{\mathbf x}$ [cf.~\eqref {eq. system anomalous hydro 0}] and $\mathbf f = 2 G_h \hat{\mathbf x}$ [cf. Eq.~\eqref{eq. system anomalous hydro 1}] yields the results for the normal and anomalous Hall effect obtained above. We stress that our theory provides microscopic expressions for $\mathbf u_a$ and $\mathbf f$ consistently taking into account all relevant spin-orbit interaction induced contributions: side-jump, anomalous velocity, and skew-scattering.}

{In the considered axially-symmetric model of electron spectrum, the off-diagonal components of the viscosity tensor appear in the presence of magnetic field and spin polarization and obey the Onsager relation $\eta_{xy} = -\eta_{yx}$. On the microscopic level the Lorentz force results in the off-diagonal contributions to the viscosity tensor (Hall viscosity)
\begin{equation}
    \label{eta:xy:ee:Lorentz}
    \eta_{xy} = -\eta_{yx} = -2\omega_c\tau_{ee}\eta, \quad |\omega_c| \tau_{ee} \ll 1.
\end{equation}
These components provide an additive contribution to the normal Hall effect [Eq.~\eqref{Hall:normal hydro}], but they can  be safely neglected in hydrodynamic regime because of the small parameter $l_{ee}/w$: The Hall viscosity manifests itself only due to $\Delta \mathbf u \ne 0$ compared to the Lorentz force action which is possible in homogeneous flow. In the presence of spin polarization the anomalous, i.e., unrelated to the Lorentz force, contributions to $\eta_{xy} = -\eta_{yx}$ appear owing to the spin-orbit interaction
\begin{equation}
    \label{eta:xy:odd}
    \eta_{xy}^{so} = -\eta_{yx}^{so} \propto P_s,
\end{equation}
where $P_s$ is the electron spin polarization degree introduced in Eq.~\eqref{Ps}. 
Such contributions for the impurity scattering were calculated and discussed in Refs.~\cite{PhysRevB.96.020401,PhysRevResearch.3.033075}. However, these terms provide negligible contribution to the anomalous Hall field $\mathbf E_H$ compared to the contributions of anomalous velocity $\mathbf u_a$ and force $\mathbf f$, since they require additional spatial gradients of velocity and, hence, contain a small factor $l_{ee}/w$.}

{However, electron-electron collisions can also contribute to the off-diagonal terms of viscosity tensor -- odd-viscosity~\cite{Avron:1998aa,PhysRevFluids.2.094101,Fruchart:2023aa} -- as
\begin{equation}
    \label{eta:xy:odd:ee}
    \eta_{xy}^{ee} = -\eta_{yx}^{ee} = \Upsilon P_s \eta,
\end{equation}
where $\Upsilon$ is a dimensionless constant related to the details of scattering (detailed calculations of $\Upsilon$ for two-dimensional electrons are beyond the present work and will be reported elsewhere). The resulting Hall field produced by odd viscosity reads
\begin{equation}
\label{Hall:odd}
E_H^{ee} = -\frac{m\eta_{xy}^{ee}}{e}\Delta V_y,
\end{equation}
where, as follows from Eq.~\eqref{eq: delta f hydro} 
\begin{equation}
V_y = \frac{eE_y}{2m\eta} \left[\left(\frac{w}{2}\right)^{2}-x^{2}\right].
\end{equation}
It results in a constant anomalous Hall field and linear voltage across the channel. We stress that the contribution~\eqref{Hall:odd} to the anomalous Hall field is inherently related to the electron-electron interactions and is present even if impurity and phonon scattering is totally inefficient, $l=\infty$.}

\section{Conclusion}\label{sec:concl}

We have developed a microscopic theory of the anomalous Hall effect in ultraclean two-dimensional electronic channels, where the transport mean free paths exceeds the width of the conducting channel. All relevant contributions to the anomalous Hall electric field have been taken into account: the anomalous velocity, the side-jump, and the skew scattering. We have studied two main regimes of the electron transport in the channel: the ballistic one where the electron mean free path (both due to the disorder scattering and electron-electron collisions) exceeds the channels width and the hydrodynamic one where the electron-electron mean free path is much smaller than the channels width. In the latter regimes we have identified two qualitatively different cases of weak and moderate magnetic fields where the Zeeman splitting is smaller or larger than the thermal energy. In all relevant cases the analytical expressions for the Hall field have been derived. 

{For the ballistic regime we have demonstrated a universal relation between the Hall voltage and spin accumulation due to the spin Hall effect. This universal relation breaks in the hydrodynamic regime where the Hall voltage and field are sensitive to the strength of electron-electron collisions and also in the deeply diffusive (Ohmic) regime where spin relaxation processes become important in the case of the spin Hall effect.}

{In the hydrodynamic regime the anomalous Hall effect is determined both by electron-collisions and momentum relaxation processes due to impurity scattering. An increase in magnetic field suppresses electron-electron collisions with opposite spins changing the spin distribution and, accordingly, the anomalous Hall field and voltage. We have shown that in the moderate magnetic field case, where the opposite spin collisions are suppressed, the anomalous Hall field caused by the skew scattering is inhomogeneous and corresponding Hall voltage is a nonlinear function of coordinate.}

{Within general hydrodynamic framework we have also discussed the effect of odd viscosity due to electron-electron collisions and shown that in the case where impurity and phonon scattering is absent, the anomalous Hall effect is solely driven by the odd viscosity.}

The anomalous contribution to the Hall field and voltage are smaller than the normal ones. The anomalous Hall effect can be significantly enhanced in diluted magnetic semiconductors where the Zeeman splitting of the charge carriers is controlled by the spin polarization of magnetic ions. In that case the anomalous contributions can be extracted owing to their characteristic magnetic field dependence -- strong linear growth followed by saturation -- and high sensitivity to the temperature. Another possible way to detect the anomalous Hall effect is to use the electron spin resonance technique: By depolarization of the electron spins via additional weak alternating electromagnetic field one can extract the spin-related -- anomalous -- contribution to the Hall field and voltage from the total one measured in conventional Hall setup.

To conclude, sensitivity of the anomalous Hall effect to (i) spin-orbit interaction and band topology, (ii) transport regime, and (iii) electron-electron interactions opens up avenues to in-depth experimental studies of these key effects in ultraclean two-dimensional electronic systems. Comparison of the developed theory and future experiments allow one to  access intricate phenomena like odd viscosity and extract important parameters governing electron transport.

\section*{Acknowledgements}

This work has been supported by the RSF project 22-12-00211. We are grateful to P.S. Alekseev and M.A. Semina for valuable comments.

%\bibliography{AHEultra.bib}


\end{document}